\documentstyle [12pt,axodraw] {article}

\catcode`@=12
\begin{document}
\thispagestyle{empty}
\begin{flushright} UCRHEP-T262\\September 1999\
\end{flushright}
\vskip 0.5in
\begin{center}
{\Large \bf Permutation Symmetry for Neutrino\\
and Charged-Lepton Mass Matrices\\}
\vskip 1.5in
{\bf Ernest Ma\\}
\vskip 0.1in
{\sl Physics Department, Univeristy of California,\\}
{\sl Riverside, CA 92521, USA\\}
\end{center}
\vskip 2.0in
\begin{abstract}\
The permutation symmetry $S_3$ is appplied to obtain two equal Majorana 
neutrino masses, while maintaining three different charged-lepton masses and 
suppressing neutrinoless double beta decay.  The resulting radiative 
splitting of the two neutrinos is shown to be suitable for solar neutrino 
vacuum oscillations.
\end{abstract}
\newpage
\baselineskip 24pt

\section{Introduction}

There are three known charged leptons, $e$, $\mu$, $\tau$, with very different 
masses:
\begin{equation}
m_e = 0.511~{\rm MeV}, ~~~ m_\mu = 105.66~{\rm MeV}, ~~~ m_\tau = 
1777~{\rm MeV}.
\end{equation}
Their accompanying neutrinos, $\nu_e$, $\nu_\mu$, $\nu_\tau$, are not 
necessarily mass eigenstates.  In general,
\begin{equation}
\nu_\alpha = \sum_{i=1}^3 U_{\alpha i} \nu_i,
\end{equation}
where $\alpha = e, \mu, \tau$ and $\nu_i$ are mass eigenstates.
\begin{equation}
U = \left[ \begin{array} {c@{\quad}c@{\quad}c} \cos \theta & \sin \theta & 0 
\\ -\sin \theta/\sqrt 2 & \cos \theta/\sqrt 2 & -1/\sqrt 2 \\ -\sin \theta 
/\sqrt 2 & \cos \theta/\sqrt 2 & 1/\sqrt 2 \end{array} \right]
\end{equation}
is a typical mixing matrix which allows one to understand the recent 
atmospheric neutrino data\cite{1} and the long-standing solar neutrino 
deficit\cite{2} in terms of neutrino oscillations.  The form of $U$ in Eq.~(3) 
has been advocated by many authors\cite{3}.  It has the virtue of maximal 
mixing between $\nu_\mu$ and $\nu_\tau$ which agrees well with the 
atmospheric data, and it allows the solar data to be interpreted with 
either the small-angle or large-angle matter-enhanced neutrino-oscillation 
solution\cite{4}, or the necessarily large-angle vacuum solution.

The masses $m_{1,2,3}$ are now subject to the conditions that
\begin{equation}
\Delta m^2_{13} \simeq \Delta m^2_{23} \sim 10^{-3}~{\rm eV}^2
\end{equation}
for atmospheric neutrino oscillations, and
\begin{equation}
\Delta m^2_{12} \sim 10^{-5} ~~{\rm or}~~ 10^{-10}~{\rm eV}^2
\end{equation}
for solar matter-enhanced or vacuum neutrino oscillations, where $\Delta 
m^2_{ij} \equiv m^2_i - m^2_j$.  This allows for the intriguing possibility 
that neutrino masses are degenerate\cite{5} with very small splittings. 
However, since the charged-lepton masses break this degeneracy, there must 
be radiative corrections\cite{6} which may or may not be consistent with 
the actual phenomenological solutions desired, especially so if a value of 
10$^{-10}$ eV$^2$ for $\Delta m^2_{12}$ is to be maintained.

In this paper, a detailed model of lepton mass matrices is presented, 
based on the permutation symmetry $S_3$.  It is the outgrowth of a previous 
proposal\cite{7} which shows how $\Delta m_{12}^2$ of order $10^{-10}$ 
eV$^2$ for solar neutrino vacuum oscillations can be obtained as a 
second-order perturbation of a two-fold degenerate neutrino mass matrix, 
resulting in a successful formula relating atmospheric and solar neutrino 
oscillations.  In Sec.~2 the model is described and it is shown how the 
breaking of $S_3$ together with the electroweak gauge symmetry allows the 
charged-lepton masses to be all different while maintaining a two-fold 
degeneracy in the meutrino mass matrix ${\cal M}_\nu$ at tree level. 
In Sec.~3 radiative corrections to ${\cal M}_\nu$ are derived in terms of the 
parameters of the charged-lepton mass matrix ${\cal M}_l$.  The latter is 
chosen such that neutrinoless double beta decay\cite{8} is absent at tree 
level. In Sec.~4 the resulting phenomenon of lepton flavor nonconservation 
beyond that of neutrino oscillations is discussed in the context of other 
processes involving charged leptons.  Finally in Sec.~5 there are some 
concluding remarks.

\section{Lepton mass matrices under $S_3$}

Let the three families of leptons be denoted by $(\nu_i, l_i)_L$ and 
$l^c_{iL}$, $i = 1,2,3$.  In this convention, $l_{iL} l^c_{jL}$ is a Dirac 
mass term for the charged leptons (instead of the usual $\bar l_{iL} l_{jR}$) 
and $\nu_i \nu_j$ is a Majorana mass term for the neutrinos.  Consider the 
discrete permutation symmetry $S_3$.  Its irreducible representations are 
\underline 2, \underline 1, and \underline 1$'$, with the following 
multiplication rules:  \underline 2 $\times$ \underline 2 = \underline 2 + 
\underline 1 + \underline 1$'$ and \underline 1$'$ $\times$ \underline 1$'$ 
= \underline 1.  Under $S_3$, let
\begin{equation}
\left[ \left( \begin{array} {c} \nu_1 \\ l_1 \end{array} \right)_L, \left( 
\begin{array} {c} \nu_2 \\ l_2 \end{array} \right)_L \right] \sim 2, ~~~ 
\left( \begin{array} {c} \nu_3 \\ l_3 \end{array} \right)_L \sim 1, ~~~ 
[ l^c_{1L}, l^c_{2L} ] \sim 2, ~~~ l^c_{3L} \sim 1,
\end{equation}
The Higgs sector of this model consists of three doublets $\Phi_i = (\phi^0_i, 
\phi^-_i), i = 1,2,3$, and one triplet $\xi = (\xi^{++}, \xi^+, \xi^0)$. 
Under $S_3$, let
\begin{equation}
(\Phi_1, \Phi_2) \sim 2, ~~~ \Phi_3 \sim 1, ~~~ \xi \sim 1.
\end{equation}
Neutrinos couple to $\xi$ according to
\begin{equation}
f_{ij}[\xi^0 \nu_i \nu_j + \xi^+ (\nu_i l_j + l_i \nu_j)/\sqrt 2 + \xi^{++} 
l_i l_j] + h.c.,
\end{equation}
where $f_{ij}$ is restricted by $S_3$ to have the form
\begin{equation}
f = \left( \begin{array}{c@{\quad}c@{\quad}c} 0 & f_0 & 0 \\ 
f_0 & 0 & 0 \\ 0 & 0 & f_3 \end{array} \right).
\end{equation}
As shown recently\cite{9,10}, this is an equally natural way to obtain small 
Majorana neutrino masses as the canonical seesaw mechanism\cite{11}, because 
the vacuum expectation value $\langle \xi^0 \rangle = u$ is inversely 
proportional to $m_\xi^2$.  Let $m_0 = 2 f_0 u$ and $m_3 = 2 f_3 u$, then 
the Majorana neutrino mass matrix spanning $\nu_{1,2,3}$ is given by
\begin{equation}
{\cal M}_\nu = \left( \begin{array}{c@{\quad}c@{\quad}c} 0 & m_0 & 0 \\ 
m_0 & 0 & 0 \\ 0 & 0 & m_3 \end{array} \right).
\end{equation}
The eigenvalues of ${\cal M}_\nu$ are $-m_0$, $m_0$, and $m_3$.  [A negative 
mass here means that the corresponding Majorana neutrino is odd under $CP$ 
after a $\gamma_5$ rotation to remove the minus sign.]  Hence there is an 
effective two-fold degeneracy in the $\nu_1 - \nu_2$ sector, and it 
corresponds to an additional global symmetry, i.e. $L_1 - L_2$, if the 
charged-lepton mass matrix ${\cal M}_l$ is diagonal in the same basis.

There are five Yukawa interaction terms of the charged leptons with the 
Higgs doublets which are invariant under $S_3$, i.e.
\begin{eqnarray}
&& h_1 [l_1 l^c_1 \phi^0_1 + l_2 l^c_2 \phi^0_2 - \nu_1 l^c_1 \phi^-_1 - 
\nu_2 l^c_2 \phi^-_2] \nonumber \\ 
&+& h_2 [(l_1 l^c_2 + l_2 l^c_1) \phi^0_3 - (\nu_1 l^c_2 + \nu_2 l^c_1) 
\phi^-_3] \nonumber \\ 
&+& h_3 [(l_1 \phi^0_2 + l_2 \phi^0_1) l^c_3 - (\nu_1 \phi^-_2 + \nu_2 
\phi^-_1) l^c_3] \nonumber \\ 
&+& h_4 [l_3 (l^c_1 \phi^0_2 + l^c_2 \phi^0_1) - \nu_3 (l^c_1 \phi^-_2 + 
l^c_2 \phi^-_1)] \nonumber \\ 
&+& h_5 [l_3 l^c_3 \phi^0_3 - \nu_3 l^c_3 \phi^-_3] + h.c.
\end{eqnarray}
As $\phi^0_{1,2,3}$ acquire vacuum expectation values $v_{1,2,3}$, the $3 
\times 3$ mass matrix linking $l_{1,2,3}$ to $l^c_{1,2,3}$ is given by
\begin{equation}
{\cal M}_l = \left[ \begin{array} {c@{\quad}c@{\quad}c} h_1 v_1 & h_2 v_3 & 
h_3 v_2 \\ h_2 v_3 & h_1 v_2 & h_3 v_1 \\ h_4 v_2 & h_4 v_1 & h_5 v_3 
\end{array} \right].
\end{equation}
The scalar potential of this model is assumed to respect $S_3$ only in its 
dimension-four terms, i.e. $S_3$ is broken \underline {softly} by its 
dimension-two and dimension-three terms:
\begin{eqnarray}
&& \sum_{i,j=1}^3 m_{ij}^2 (\bar \phi^0_i \phi^0_j + \phi^+_i \phi^-_j) + 
m_\xi^2 (\xi^{--} \xi^{++} + \xi^- \xi^+ + \bar \xi^0 \xi^0) \nonumber \\ 
&& + \sum_{i,j=1}^3 \mu_{ij} \left[ \xi^{++} \phi^-_i \phi^-_j + {1 \over 
\sqrt 2} \xi^+ (\phi^0_i \phi^-_j + \phi^-_i \phi^0_j) + \xi^0 \phi^0 \phi^0 
\right] + h.c.
\end{eqnarray}
This allows ${\cal M}_l$ to break $S_3$ with $v_2 \neq v_1$.  In fact, the 
limits $h_2 = 0$ and $v_2 = 0$ are assumed here so that ${\cal M}_l$ becomes 
of the form
\begin{equation}
{\cal M}_l = \left( \begin{array}{c@{\quad}c@{\quad}c} m_e & 0 & 0 \\ 
0 & 0 & a \\ 0 & b & d \end{array} \right),
\end{equation}
with $l_1$ identified as the electron, and
\begin{equation}
m^2_{\tau, \mu} = {1 \over 2} (d^2 + a^2 + b^2) \pm {1 \over 2} \sqrt 
{d^2 + (a + b)^2} \sqrt {d^2 + (a - b)^2}.
\end{equation}

From Eqs.~(10) and (14), the eigenstates of ${\cal M}_\nu$ are easily read 
off:
\begin{equation}
\nu_1 = {1 \over \sqrt 2} (\nu_e - c \nu_\mu - s \nu_\tau), ~~~ 
\nu_2 = {1 \over \sqrt 2} (\nu_e + c \nu_\mu + s \nu_\tau), ~~~ 
\nu_3 = -s \nu_\mu + c \nu_\tau,
\end{equation}
where $c = \cos \theta_L$ and $s = \sin \theta_L$ are determined by 
the $\mu_L - \tau_L$ sector of Eq.~(14).  This means that $\nu_e$ mixes 
maximally with $c \nu_\mu + s \nu_\tau$, i.e. $\theta = \pi/4$ in Eq.~(3).  
If $c = s = 1/\sqrt 2$ is also assumed in the above (corresponding to 
$a^2 = b^2 + d^2$), the so-called bimaximal form of neutrino oscillations is 
obtained.  In that case,
\begin{equation}
a^2 = {1 \over 2} (m_\tau^2 + m_\mu^2), ~~~ b^2 = {2 m_\tau^2 m_\mu^2 \over 
m_\tau^2 + m_\mu^2}, ~~~ d^2 = {(m_\tau^2 - m_\mu^2)^2 \over 2(m_\tau^2 + 
m_\mu^2)}.
\end{equation}

\section{Radiative corrections to neutrino mass degeneracy}

To discuss radiative corrections to ${\cal M}_\nu$, consider first the case 
of keeping only one Higgs doublet $\Phi$ and one one Higgs triplet $\xi$. 
In this scenario, $S_3$ is explicitly broken by the Yukawa interactions of 
the charged leptons.  Consequently, there is an arbitrariness in choosing 
the mass scale at which $S_3$ is assumed to be exact.  The most natural 
choice in the present context is of course $m_\xi$, hence there are two 
contributions to the radiatively corrected ${\cal M}_\nu$.  One is a finite 
correction to the mass matrix, as shown in Figure 1.  The other is a 
renormalization of the coupling matrix from the shift in mass scale from 
$m_\xi$ to $m_W$.  This was the specific case presented in Ref.~[7].

Here the situation is different in two ways.  First, $S_3$ is a good symmetry 
as far as the Yukawa couplings are concerned.  Second, it is softly broken 
only at the electroweak energy scale.  Hence there is no $S_3$ breaking 
contribution from the renormalization of the coupling matrix.  As for the 
finite correction to the mass matrix, because of the approximations $h_2 = 0$ 
and $v_2 = 0$, there are now only two contributions: $\nu_1 \nu_3 
\bar \phi^0_1 \bar \phi^0_3$ and $\nu_3 \nu_3 \bar \phi^0_3 \bar \phi^0_3$, 
as shown in Figure 1.  Assuming that $\mu_{33}$ dominates among all the 
$\mu_{ij}$'s in Eq.~(13) so that\cite{9} $u \simeq -\mu_{33} v_3^2/m_\xi^2$, 
the mass matrix ${\cal M}_\nu$ is now corrected to read
\begin{equation}
{\cal M}_\nu = \left( \begin{array}{c@{\quad}c@{\quad}c} 0 & m_0 & 
adI m_0 \\ m_0 & 0 & 0 \\ adI m_0 & 0 & (1+2 d^2 I) m_3
\end{array} \right).
\end{equation}
The integral $I$ is given by\cite{7}
\begin{equation}
I = {G_F \over 4 \pi^2 \sqrt 2 \sin^2 \beta} \ln {m_\xi^2 \over m_W^2},
\end{equation}
where $\sin^2 \beta = v_3^2/(v_1^2 + v_3^2)$.  The two-fold degeneracy of the 
$\nu_1 - \nu_2$ sector is then lifted, with the following mass eigenvalues:
\begin{equation}
-m_0 - {a^2 d^2 I^2 m_0^2 \over 2 (m_0+m_3)}, ~~~ 
m_0 + {a^2 d^2 I^2 m_0^2 \over 2 (m_0-m_3)},
\end{equation}
where $ a d I << (m_0 - m_3)/(m_0 + m_3)$ has been used, being justified 
numerically. Hence their mass-squared difference is
\begin{equation}
\Delta m^2 \simeq a^2 d^2 I^2 m_0^3 \left[ { 1 \over m_0-m_3} - 
{ 1 \over m_0+m_3} \right] \simeq {2 a^2 d^2 I^2 m_\nu^4 \over m_0^2 
-m_3^2},
\end{equation}
where $m_\nu \simeq m_0 \simeq m_3$ has been used.  Identifying this with 
solar neutrino vacuum oscillations then yields
\begin{equation}
{(\Delta m^2)_{sol} (\Delta m^2)_{atm} \over m_\nu^4} = 2 a^2 d^2 I^2 = 
{2.16 \times 10^{-13} \over \sin^4 \beta} \left( \ln {m_\xi^2 \over m_W^2} 
\right)^2,
\end{equation}
where $a = 1259$ MeV and $d = 1250$ MeV from Eq.~(17) have been used. In the 
above, bimaximal mixing (i.e. $\sin^2 2 \theta_{sol} = \sin^2 2 \theta_{atm} 
= 1$) has been assumed.  For $(\Delta m^2)_{sol} \sim 4 \times 10^{-10}$ 
eV$^2$ in the case of vacuum oscillations and $(\Delta m^2)_{atm} \sim 4 
\times 10^{-3}$ eV$^2$, this would require
\begin{equation}
{m_\nu^4 \over \sin^4 \beta} \left( \ln {m_\xi^2 \over m_W^2} \right)^2 
\sim 7.4 ~{\rm eV}^4,
\end{equation}
which gives the bound $m_\nu < 0.6$ eV for $m_\xi > 1$ TeV and $\sin^2 
\beta < 0.7$.  It is interesting to note that this same numerical limit 
in the case of three nearly mass-degenerate neutrinos was recently 
obtained\cite{12} from the consideration of cosmic structure formation 
in the light of the latest astronomical observations. 

The choice of Eq.~(14) in conjunction with Eq.~(10) means that neutrinoless 
double beta decay\cite{8} is absent to lowest order.  It also eliminates 
any one-loop correction to the diagonal entries of ${\cal M}_\nu$ in the 
$\nu_1 - \nu_2$ sector.  This allows the mass splitting to be quadratic 
(as opposed to linear) in $I$ as shown in Eq.~(20), which is crucial for 
obtaining the very small phenomenological value of $(\Delta m^2)_{sol}$ 
for vacuum oscillations.

\section{Lepton flavor nonconservation}

Both lepton number and lepton flavor are not conserved in this model.  Whereas 
lepton number nonconservation originates from the heavy Higgs triplet $\xi$ 
and manifests itself at low energy only through the very small Majorana 
neutrino masses, lepton flavor nonconservation originates from the much less 
heavy Higgs doublets which are presumably in the 100 GeV mass range.  On 
the other hand, the $h_i$'s of Eq.~(11) are suppressed relative to the gauge 
couplings because they are related to ${\cal M}_l$ as shown in Eqs.~(12) and 
(14).

Using Eqs.~(14) and (17), it is easily shown that
\begin{equation}
\left( \begin{array} {c} l_1 \\ l_2 \\ l_3 \end{array} \right) = \left( 
\begin{array} {c@{\quad}c@{\quad}c} 1 & 0 & 0 \\ 0 & 1/\sqrt 2 & -1/\sqrt 2 \\ 
0 & 1/\sqrt 2 & 1/\sqrt 2 \end{array} \right) \left( \begin{array} {c} e \\ 
\mu \\ \tau \end{array} \right),
\end{equation}
and
\begin{equation}
\left( \begin{array} {c} l_1^c \\ l_2^c \\ l_3^c \end{array} \right) = \left( 
\begin{array} {c@{\quad}c@{\quad}c} 1 & 0 & 0 \\ 0 & \cos \theta_R & 
\sin \theta_R \\ 0 & -\sin \theta_R & \cos \theta_R \end{array} \right) 
\left( \begin{array} {c} e^c \\ \mu^c \\ \tau^c \end{array} \right),
\end{equation}
where
\begin{equation}
\tan \theta_R = {m_\mu \over m_\tau} \left( {m_\tau^2 - m_\mu^2 \over m_\tau^2 
+ m_\mu^2} \right).
\end{equation}
The interactions of $\phi^0_{1,2,3}$ are then given by
\begin{eqnarray}
\phi^0_1 \left[ h_1 e e^c + {-h_3 s_R + h_4 c_R \over \sqrt 2} \mu \mu^c  
+ {-h_3 c_R + h_4 s_R \over \sqrt 2} \tau \tau^c + {h_3 c_R + h_4 s_R 
\over \sqrt 2} \mu \tau^c + {h_3 s_R + h_4 c_R \over \sqrt 2} \tau \mu^c 
\right] + \nonumber \\ 
\phi^0_2 \left[ {h_1 c_R \over \sqrt 2} \mu \mu^c - {h_1 s_R \over 
\sqrt 2} \tau \tau^c - h_3 s_R e \mu^c + {h_4 \over \sqrt 2} \mu e^c + 
h_3 c_R e \tau^c + {h_4 \over \sqrt 2} \tau e^c + {h_1 s_R \over \sqrt 2} 
\mu \tau^c - {h_1 c_R \over \sqrt 2} \tau \mu^c \right] + \nonumber \\ 
\phi^0_3 \left[ -{h_5 s_R \over \sqrt 2} \mu \mu^c + {h_5 c_R \over 
\sqrt 2} \tau \tau^c + h_1 c_R e \mu^c + {h_1 \over \sqrt 2} \mu e^c + h_1 
s_R e \tau^c - {h_1 \over \sqrt 2} \tau e^c + {h_5 c_R \over \sqrt 2} \mu 
\tau^c - {h_5 s_R \over \sqrt 2} \tau \mu^c \right], \nonumber \\
\end{eqnarray}
where $s_R = \sin \theta_R$ and $c_R = \cos \theta_R$.  In the above,
\begin{equation}
h_1 = {m_e \over v_1}, ~~ h_3 \simeq {m_\tau \over v_1 \sqrt 2}, ~~ h_4 
\simeq {\sqrt 2 m_\mu \over v_1}, ~~ h_5 \simeq {m_\tau \over v_3 \sqrt 2}, 
~~ s_R \simeq {m_\mu \over m_\tau}, ~~ c_R \simeq 1.
\end{equation}

Consequently, the most prominent rare decays are $\tau^- \to e^- e^- \mu^+$, 
$\tau^- \to e^- \mu^- \mu^+$, and $\tau^- \to \mu^- \mu^- \mu^+$, with 
branching fractions
\begin{equation}
B(\tau^- \to e^- e^- \mu^+) \simeq 2 B(\tau^- \to e^- \mu^- \mu^+) \simeq 
{m_\mu^2 m_\tau^2 \over 8 \cos^4 \beta m^4_{\phi^0_2}},
\end{equation}
and
\begin{equation}
B(\tau^- \to \mu^- \mu^- \mu^+) \simeq {m_\mu^2 m_\tau^2 \over 16} \left( 
{1 \over \cos^2 \beta m^2_{\phi^0_1}} - {1 \over \sin^2 \beta m^2_{\phi^0_3}} 
\right)^2.
\end{equation}
With the scalar masses of order 100 GeV, these branching fractions are of 
order $10^{-10}$, much below the present experimental upper limits\cite{13} 
which are of order $10^{-6}$.  Note that $\mu \to e e e$ is suppressed even 
more strongly in this model because its amplitude is proportional to 
$m_e m_\mu$.

Whereas low-energy tests of this model are limited to neutrino masses and 
oscillations, dramatic effects are predicted at high energies.  The 
production of $\phi^0_{1,2,3}$ at future colliders would yield very clear 
signals from decays such as $\phi^0_2 \to \tau^- e^+$ and $\phi^0_{1,3} \to 
\tau^- \mu^+$.

\section{Concluding remarks}

To understand the present experimental data on atmospheric and solar 
neutrinos, a model of neutrino and charged-lepton mass matrices based on 
the permutation symmetry $S_3$ has been proposed.  It has a two-fold 
degeneracy in the neutrino mass matrix which is broken radiatively, and 
allows for a very small mass splitting, suitable for solar neutrino 
vacuum oscillations, as given by Eq.~(22).  The $S_3$ symmetry is maintained 
in the scalar sector by three Higgs doublets which determine the 
charged-lepton mass matrix, whereas the neutrinos obtain naturally small 
Majorana masses from their couplings to a heavy Higgs triplet.  Lepton 
flavor nonconservation at low energy is suppressed by the small charged-lepton 
masses.  This model may be tested at high energy with the production and 
decay of its scalar doublets.
\vskip 0.3in
\begin{center} {ACKNOWLEDGEMENT}
\end{center}

This work was supported in part by the U.~S.~Department of Energy under 
Grant No.~DE-FG03-94ER40837.

\bibliographystyle{unsrt}

\begin{thebibliography}{99}
\bibitem{1} Y. Fukuda {\it et al.}, Phys. Lett. {\bf B433}, 9 (1998); 
{\bf B436}, 33 (1998); Phys. Rev. Lett. {\bf 81}, 1562 (1998); {\bf 82}, 2644 
(1999); {\bf 82}, 5194 (1999); hep-ex/9908049.
\bibitem{2} R. Davis, Prog. Part. Nucl. Phys. {\bf 32}, 13 (1994);  P. 
Anselmann {\it et al.}, Phys. Lett. {\bf B357}, 237 (1995); {\bf B361}, 235 
(1996); J. N. Abdurashitov {\it et al.}, Phys. Lett. {\bf B328}, 234 (1994); 
astro-ph/9907131; 
Y. Fukuda {\it et al.}, Phys. Rev. Lett. {\bf 77}, 1683 (1996); {\bf 81}, 
1158 (1998); {\bf 82}, 1810 (1999); {\bf 82}, 2430 (1999).
\bibitem{3} C. H. Albright, K. S. Babu, and S. M. Barr, Phys. Rev. Lett. 
{\bf 81}, 1167 (1998); 
V. Barger, S. Pakvasa, T. J. Weiler, and K. Whisnant, Phys. Lett. 
{\bf B437}, 107 (1998);
S. F. King, Phys. Lett. {\bf B439}, 350 (1998); hep-ph/9904210; 
B. C. Allanach, Phys. Lett. {\bf B450}, 182 (1999); 
J. K. Elwood, N. Irges, and P. Ramond, Phys. Rev. Lett. {\bf 81}, 5064 (1998); 
R. Barbieri, L. J. Hall, D. Smith, A. Strumia, and N. Weiner, JHEP {\bf 9812}, 
017 (1998); 
A. J. Baltz, A. S. Goldhaber, and M. Goldhaber, Phys. Rev. Lett. {\bf 81}, 
5730 (1998);
G. Altarelli and F. Feruglio, Phys. Lett. {\bf B439}, 112 (1998); JHEP 
{\bf 9811}, 021 (1998); Phys. Lett. {\bf B451}, 388 (1999); 
M. Jezabek and Y. Sumino, Phys. Lett. {\bf B440}, 327 (1998); {\bf B457}, 139 
(1999); 
H. Fritzsch and Z. Xing, Phys. Lett. {\bf B440}, 313 (1998); 
R. N. Mohapatra and S. Nussinov, Phys. Lett. {\bf B441}, 299 (1998); 
E. Ma, Phys. Lett. {\bf B442}, 238 (1998); 
Y. Nomura and T. Yanagida, Phys. Rev. {\bf D59}, 017303 (1999); 
M. Tanimoto, Phys. Rev. {\bf D59}, 017304 (1999); Phys. Lett. {\bf B456}, 220 
(1999);
N. Haba, Phys. Rev. {\bf D59}, 035011 (1999); 
K. Oda, E. Takasugi, M. Tanaka, and M. Yoshimura, Phys. Rev. {\bf D59}, 
055001 (1999); 
J. Ellis, G. K. Leontaris, S. Lola, and D. V. Nanopoulos, Eur. Phys. J. 
{\bf C9}, 389 (1999); 
E. Ma, D. P. Roy, and U. Sarkar, Phys. Lett. {\bf B444}, 391 (1998);  
S. Davidson and S. F. King, Phys. Lett. {\bf B445}, 191 (1998); 
R. Barbieri, L. J. Hall, and A. Strumia, Phys. Lett. {\bf B445}, 407 (1999); 
Y. Grossman, Y. Nir, and Y. Shadmi, JHEP {\bf 9810}, 007 (1998);  
S. L. Glashow, P. J. Kernan, and L. M. Krauss, Phys. Lett. {\bf B445}, 412 
(1999);  
A. S. Joshipura and S. D. Rindani, hep-ph/9811252;  hep-ph/9906390; 
C. Jarlskog, M. Matsuda, S. Skadhauge, and M. Tanimoto, Phys. Lett. 
{\bf B449}, 240 (1999);
E. Malkawi, hep-ph/9810542;  
E. Ma and D. P. Roy, Phys. Rev. {\bf D59}, 097702 (1999);  
M. Fukugita, M. Tanimoto, and T. Yanagida, Phys. Rev. {\bf D59}, 113016 
(1999);  
L. J. Hall and N. Weiner, Phys. Rev. {\bf D60}, 033005 (1999); 
Z. Berezhiani and A. Rossi, JHEP {\bf 9903}, 002 (1999); 
K. S. Babu, J. C. Pati, and F. Wilczek, hep-ph/9812538; 
S. Lola and G. G. Ross, Nucl. Phys. {\bf B553}, 81 (1999);
Q. Shafi and Z. Tavartkiladze, Phys. Lett. {\bf B451}, 129 (1999);
Y. Nomura and T. Sugimoto, hep-ph/9903334;
T. Blazek, S. Raby, and K. Tobe, hep-ph/9903340;
R. Barbieri, P. Creminelli, and A. Romanino, hep-ph/9903460;
K. S. Babu, B. Dutta, R. N. Mohapatra, Phys. Lett. {\bf B458}, 93 (1999); 
C.-K. Chua, X.-G. He, and W.-Y. P. Hwang, hep-ph/9905340; 
B. Stech, hep-ph/9905440; 
H. B. Benaoum and S. Nasri. hep-ph/9906232; 
C. H. Albright and S. M. Barr, hep-ph/9906297; 
P. H. Frampton and S. L. Glashow, hep-ph/9906375; 
J. C. Romao, M. A. Diaz, M. Hirsch, W. Porod, and J. W. F. Valle, 
hep-ph/9907499; 
G. Altarelli, F. Feruglio, and I. Masina, hep-ph/9907532;
A. Abada and M. Losada, hep-ph/9908352.
\bibitem{4} L. Wolfenstein, Phys. Rev. {\bf D17}, 2369 (1978); S. P. 
Mikheyev and A. Yu. Smirnov, Sov. J. Nucl. Phys. {\bf 42}, 913 (1986).
\bibitem{5} D. Caldwell and R. N. Mohapatra, Phys. Rev. {\bf D48}, 3259 
(1993);  
A. S. Joshipura, Z. Phys. {\bf C64}, 31 (1994); Phys. Rev. {\bf D51}, 
1321 (1995); 
P. Bamert and C. P. Burgess, Phys. Lett. {\bf B329}, 289 (1994); 
D.-G. Lee and R. N. Mohapatra, Phys. Lett. {\bf B329}, 463 (1994); 
A. Ioannisian and J. W. F. Valle, Phys. Lett. {\bf B332}, 93 (1994); 
A. Ghosal, Phys. Lett. {\bf B398}, 315 (1997); 
A. K. Ray and S. Sarkar, Phys. Rev. {\bf D58}, 055010 (1998); 
C. D. Carone and M. Sher, Phys. Lett. {\bf B420}, 83 (1998); 
F. Vissani, hep-ph/9708483; 
H. Georgi and S. L. Glashow, hep-ph/9808293; 
U. Sarkar, Phys. Rev. {\bf D59}, 037302 (1999); 
R. N. Mohapatra and S. Nussinov, Phys. Rev. {\bf D60}, 013002 (1999); 
G. C. Branco, M. N. Rebelo, and J. I. Silva-Marcos, Phys. Rev. Lett. 
{\bf 82}, 683 (1999); hep-ph/9906368;
C. Wetterich, Phys. Lett. {\bf B451}, 397 (1999); 
Y. L. Wu, hep-ph/9810491 (Phys. Rev. {\bf D}, in press); hep-ph/9901245 (Eur. 
Phys. J. {\bf C}, in press); hep-ph/9901320 (Int. J. Mod. Phys. {\bf A}, in 
press); hep-ph/9905222; hep-ph/9906435; 
R. Barbieri, L. J. Hall, G. L. Kane, and G. G. Ross, hep-ph/9901228;
M. Tanimoto, T. Watari, and T. Yanagida, hep-ph/9904338;
A. K. Ray and S. Sarkar, hep-ph/9908294.
\bibitem{6} E. Ma, Phys. Lett. {\bf B456}, 48 (1999); {\bf B456}, 201 (1999); 
hep-ph/9907400 (J. Phys. {\bf G}, in press); hep-ph/9907503; 
J. Ellis and S. Lola, Phys. Lett. {\bf B458}, 310 (1999); 
J. A. Casas, J. R. Espinosa, A. Ibarra, and I. Navarro, hep-ph/9904395; 
hep-ph/9905381; hep-ph/9906281;
R. Barbieri, G. G. Ross, and A. Strumia, hep-ph/9906470;
N. Haba and N. Okamura, hep-ph/9906481;
N. Haba, Y. Matsui, N. Okamura, and M. Sugiura, hep-ph/9908429.
\bibitem{7} E. Ma, hep-ph/9902392 (Phys. Rev. Lett., in press).
\bibitem{8} L. Baudis {\it et al.}, Phys. Rev. Lett. {\bf 83}, 41 (1999); 
H. V. Klapdor-Kleingrothaus, hep-ex/9901021.
\bibitem{9} E. Ma and U. Sarkar, Phys. Rev. Lett. {\bf 80}, 5716 (1998). 
\bibitem{10} E. Ma, Phys. Rev. Lett. {\bf 81}, 1171 (1998).
\bibitem{11} M. Gell-Mann, P. Ramond, and R. Slansky, in {\it Supergravity}, 
edited by P. van Nieuwenhuizen and D. Z. Freedman (North-Holland, Amsterdam, 
1979), p.~315; T. Yanagida, in {\it Proceedings of the Workshop on the 
Unified Theory and the Baryon Number in the Universe}, edited by O. Sawada 
and A. Sugamoto (KEK Report No.~79-18, Tsukuba, Japan, 1979), p.~95; R. N. 
Mohapatra and G. Senjanovic, Phys. Rev. Lett. {\bf 44}, 912 (1980).
\bibitem{12} M. Fukugita, G.-C. Liu, and N. Sugiyama, hep-ph/9908450.
\bibitem{13} Particle Data Group: C. Caso {\it et al.}, Eur. Phys. J. 
{\bf C3}, 1 (1998).
\end{thebibliography}

\begin{center}
\begin{picture}(360,200)(0,0)
\ArrowLine(30,0)(90,0)
\Text(60,-10)[c]{$\nu_{1,3}$}
\ArrowLine(180,0)(90,0)
\Text(135,-10)[c]{$l_{2,3}$}
\ArrowLine(180,0)(270,0)
\Text(225,-10)[c]{$l^c_3$}
\ArrowLine(330,0)(270,0)
\Text(300,-10)[c]{$\nu_3$}
\DashArrowLine(180,0)(180,-40)6
\Text(180,-50)[c]{$\langle \phi^0_{1,3} \rangle$}
\DashArrowLine(180,57)(180,97)6
\Text(180,106)[c]{$\langle \phi^0_3 \rangle$}
\DashArrowArc(180,-45)(101,90,154)6
\Text(130,52)[c]{$\xi^+$}
\DashArrowArcn(180,-45)(101,90,26)6
\Text(240,52)[c]{$\phi^-_3$}
\end{picture}
\vskip 2.0in
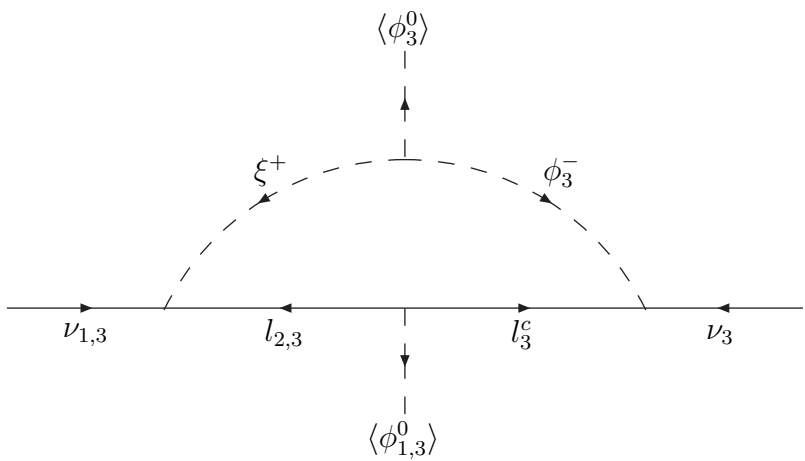
{\bf Fig.~1.} ~ One-loop radiative breaking of neutrino mass degeneracy.
\end{center}

\end{document}